# Multiple-line study of molecular gas in spiral galaxy NGC 2903

**Selçuk TOPAL**[*]

Department of Physics, Faculty of Science, Van Yüzüncü Yıl University, Van, Turkey

**Abstract:** Multiple molecular lines with radiative transfer modeling are a powerful tool to probe the physics of star-forming gas in galaxies. We investigate the gas properties in the center of spiral galaxy NGC 2903 using low-$J$ CO lines, i.e. $^{12}$CO(1-0), $^{12}$CO(2-1), $^{12}$CO(3-2), $^{13}$CO(1-0), and HCN(1-0). We apply a nonlocal thermodynamic equilibrium radiative transfer code to derive beam-averaged molecular gas properties. We use two methods (i.e. $\chi^2$ minimization and likelihood) to define the best model representing the observed line ratios best. The line ratio diagnostics suggest that CO gas in the center of NGC 2903 is thinner and the dense gas fraction is similar compared to that of spirals, starbursts, and early-type galaxies (ETGs), while the gas in the center of the galaxy is warmer than that of ETGs and colder than that of starbursts. Based on the best-fitting model results, we find that the beam-averaged gas kinetic temperature is $T_K = 20$ K, H$_2$ volume number density is $\log(n(\mathrm{H}_2)) = 4.2$ cm$^{-3}$, and CO column number density is $\log(N(\mathrm{CO})) = 19.0$ cm$^{-2}$ in the center of NGC 2903. Both methods, i.e. the line ratio diagnostics and modeling, indicate an ISM in the center of NGC 2903 having intermediate temperature and star formation activity (also supported by star formation rates), thinner CO gas with similar dense gas fraction, and higher H$_2$ volume number density compared to that of spirals, ETGs, and starbursts.

## 1. Introduction

The universe is filled with different types of galaxies with a variety of unique shapes and characteristics. In contrast to elliptical galaxies (called early-type galaxies, ETGs), spiral galaxies have a substantial amount of molecular gas, the fuel for star formation in galaxies. Spirals are thus good laboratories to study formation and evolution of stars and galaxies.

The first detection of molecules in the interstellar medium (ISM) was CH [1]. This was followed by the detections of some other molecules in the ISM, such as OH, NH$_3$, and water vapor [2–4]. Development of mm astronomy in the 1970s led us to detect carbon monoxide ($^{12}$C$^{16}$O, hereafter CO), the second most abundant molecule in the ISM [5–9]. Since molecular clouds are cold ($T$ ~10 K), it is almost impossible to detect hydrogen molecules (hereafter H$_2$), the most abundant molecule in the universe, requiring a temperature of 540 K to be excited [10]. However, CO is excited at much lower temperatures ($\approx$5.5 K) as seen in the molecular clouds and it is readily observable from the ground [11]. CO is therefore a useful indirect probe for molecular gas. In addition, high-density tracers, such as hydrogen cyanide (HCN), enable us to probe denser parts of the molecular clouds.

[*]Correspondence: selcuktopal@yyu.edu.tr

Stars are born in such cold and dense regions of molecular clouds. Since every molecule requires different physical conditions to be excited (e.g., different temperatures and densities), it is more advantageous to study such regions using multiple molecular emission lines (e.g., multiple CO lines and high-density tracers) to better reveal the physical conditions. Observation and analysis of molecular emission lines not only increase our understanding of star formation processes but also allow us to better understand the evolution of galaxies and the universe as a whole.

How do stars form? What are the effects of stellar feedbacks (e.g., strong UV radiation, stellar winds from young massive stars, and supernova explosions) on the physics of star-forming molecular gas? How are observed line ratios connected to the environmental conditions in different types of galaxies across the cosmic sea? These are some of the outstanding questions in astrophysics that we aim to focus on in this study by exploiting molecular line ratios.

We probe the physical properties of the molecular gas in the central region of NGC 2903 using multiple low-$J$ CO lines (i.e. $^{12}$CO(1-0), $^{12}$CO(2-1), $^{12}$CO(3-2), and $^{13}$CO(1-0)) and HCN(1-0), a standard high-density tracer usually bright in disc galaxies. NGC 2903 is a nearby spiral galaxy with a star formation rate (SFR) of about 2.9 $M_\odot$ $year^{-1}$, higher than that of our own galaxy, the Milky Way [13,14]. General properties of NGC 2903 are listed in Table 1. The first goal of this study is to probe the gas physical conditions in the center of NGC 2903 by applying two methods: line ratio diagnostics and a nonlocal thermodynamic equilibrium (non-LTE) radiative transfer code (RADEX [15]), which assumes large velocity gradient (LVG) approximation [16,17]. The second goal is to compare the conditions in the center of NGC 2903 to those of the center of other galaxies selected from the literature. Finally, we aim to discuss the role of stellar feedbacks (e.g., supernova explosions and existence of young OB stars nearby) in shaping the ISM in the center of NGC 2903.

Table 1. General properties of NGC 2903.

| Property | Value |
|---|---|
| Galaxy type [a] | SAB(rs)bc |
| RA (J2000) [a] | $09^h$ $32^m$ $10,1^s$ |
| Dec. (J2000) [a] | +21°30' 03'' |
| Distance [b] | 9.3 Mpc |
| Major-axis diameter [a] | 12.6 arcmin |
| Minor-axis diameter [a] | 6.0 arcmin |
| Position angle [b] | 22 degrees |
| Inclination [b] | 67.1 degrees |
| $V_\odot^a$ | 550 km/s |
| SFR [c] | 2.9 $M_\odot$ $year^{-1}$ |

References: (a): NASA Extragalactic Database; (b): [12]; (c): [13].

This paper is organized as follows. Section 2 describes the literature data obtained for spiral galaxy NGC 2903. Section 3 presents imaging and analysis of the data. Section 4 describes the radiative transfer code used to model the molecular gas properties. Section 5 presents the outcome of our line ratio diagnostics and discussions. We conclude briefly in Section 6.

## 2. Literature data

$^{12}$CO(1-0) observations of NGC 2903 were made as part of the Berkeley-Illinois-Maryland Association Survey of Nearby Galaxies (BIMA SONG) using the BIMA telescope array and National Radio Astronomy Observatory (NRAO) telescope [18]. NGC 2903 was observed in $^{12}$CO(2-1) emission using the 30-m Institut de Radioastronomie Millimétrique (IRAM) telescope as part of the HERA CO-Line Extragalactic Survey (HERACLES [19]). Additional literature data on $^{12}$CO(3-2), $^{13}$CO(1-0), and HCN(1-0) emissions were taken from the work of Mao et al. [20], Muraoka et al. [21], and Gao and Solomon [22], respectively. The beam sizes (i.e. the beam major and minor axis) for the observations of $^{12}$CO(1-0), $^{13}$CO(1-0), $^{12}$CO(2-1), $^{12}$CO(3-2), and HCN(1-0) are 6.9'' × 5.0'', 15.2'', 13.4'', 21.8'', and 28'', respectively. Please see the related papers for more details on the observations and the initial data reductions. General properties of all the data taken from the literature for NGC 2903 are listed in Table 2.

**Table 2**. General properties of the data obtained for NGC 2903.

| Transition | Rest frequency (GHz) | Telescope | Beam size ('') | Linear size* (kpc) |
|---|---|---|---|---|
| $^{12}$CO(1-0) | 115.3 | BIMA+12m | 6.9 × 5.0 | 0.3 |
| $^{12}$CO(2-1) | 230.5 | IRAM 30m | 13.4 | 0.5 |
| $^{12}$CO(3-2) | 345.8 | HHT | 21.8 | 0.8 |
| $^{13}$CO(1-0) | 110.2 | Nobeyama 45m | 15.2 | 0.6 |
| HCN(1-0) | 88.6 | IRAM 30m | 28.0 | 1.1 |

*Linear size of the central region studied at the distance of the galaxy.

## 3. Imaging and analysis

### 3.1. Imaging

To define the spatial extent of the source emission, the data cubes of both $^{12}$CO(1-0) and $^{12}$CO(2-1) emissions were first Hanning-smoothed spectrally and Gaussian-smoothed spatially with an FWHM equal to that of the beam. The smoothed cubes were clipped at the $3\sigma$ threshold (where $\sigma$ is the rms noise of the smoothed cube estimated using only channels free of emission) and then the integrated intensity maps (hereafter moment 0) were created using MIRIAD [23]. We defined the region for the contiguous source emission on these (smoothed) moment 0 maps using the Interactive Data Language (IDL) region-growing algorithm *label_region* and then created a 3D mask from these 2D maps. We applied these 3D masks to the original (unsmoothed) 3D cubes to define the spatial extent of the source emission in NGC 2903 and finally obtained the moment 0 maps and velocity maps (hereafter moment 1). Moment maps of the galaxy are shown in Figure 1.

### 3.2. Integrated intensities and total molecular gas mass at the center

Since the literature data were obtained with different telescopes, the observed beam sizes are also different. The $^{12}$CO(1-0) and $^{12}$CO(2-1) cubes were therefore convolved to the common beam size of 28'', which is the beam size of HCN(1-0) observation. We then extracted the central spectra from the (convolved) cubes using MIRIAD. The spectra of all lines detected at the center of the galaxy are shown in Figure 2. To obtain the integrated intensities at the center of the galaxy, we fit all our integrated spectra with the Gaussian function. The form of the Gaussian function that we used is:

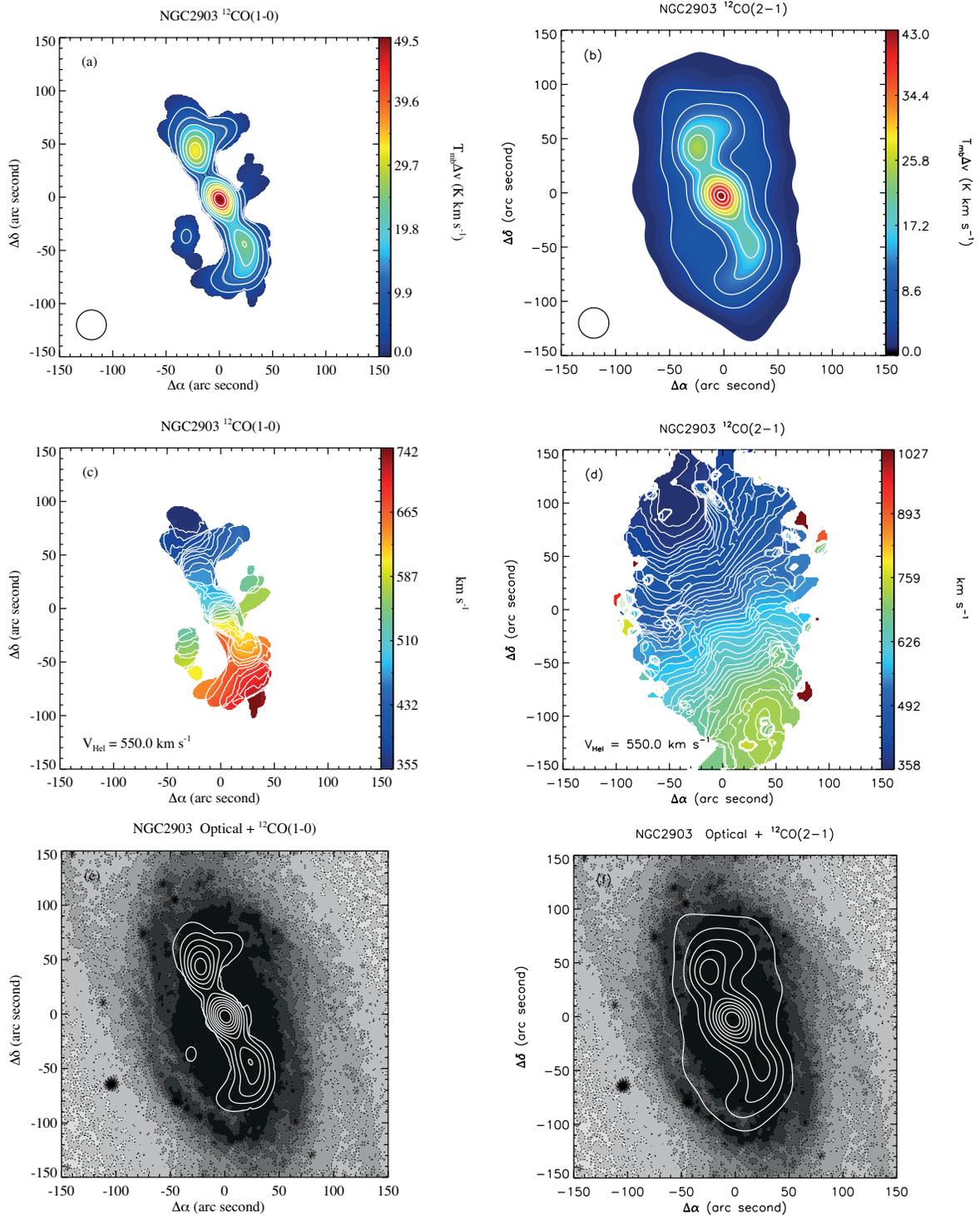

**Figure 1**. NGC 2903 moment maps. Top: moment 0 maps with overlaid isophotal contours. The small black circle at the bottom left of each panel indicates the adopted beam size of 28''. Contour levels on the moment 0 maps are from 10% to 100% of the peak intensity in steps of 10%. Middle: moment 1 maps with overlaid isovelocity contours. Contour levels on the moment 1 maps are spaced by 10 km/s. Heliocentric velocity of NGC 2903 is shown at the bottom left of each panel. Bottom: moment 0 maps of NGC 2903 are overlaid on the optical image of the galaxy (gray-scale) taken from the Sloan Digital Sky Survey. North is up and east to the left in all images.

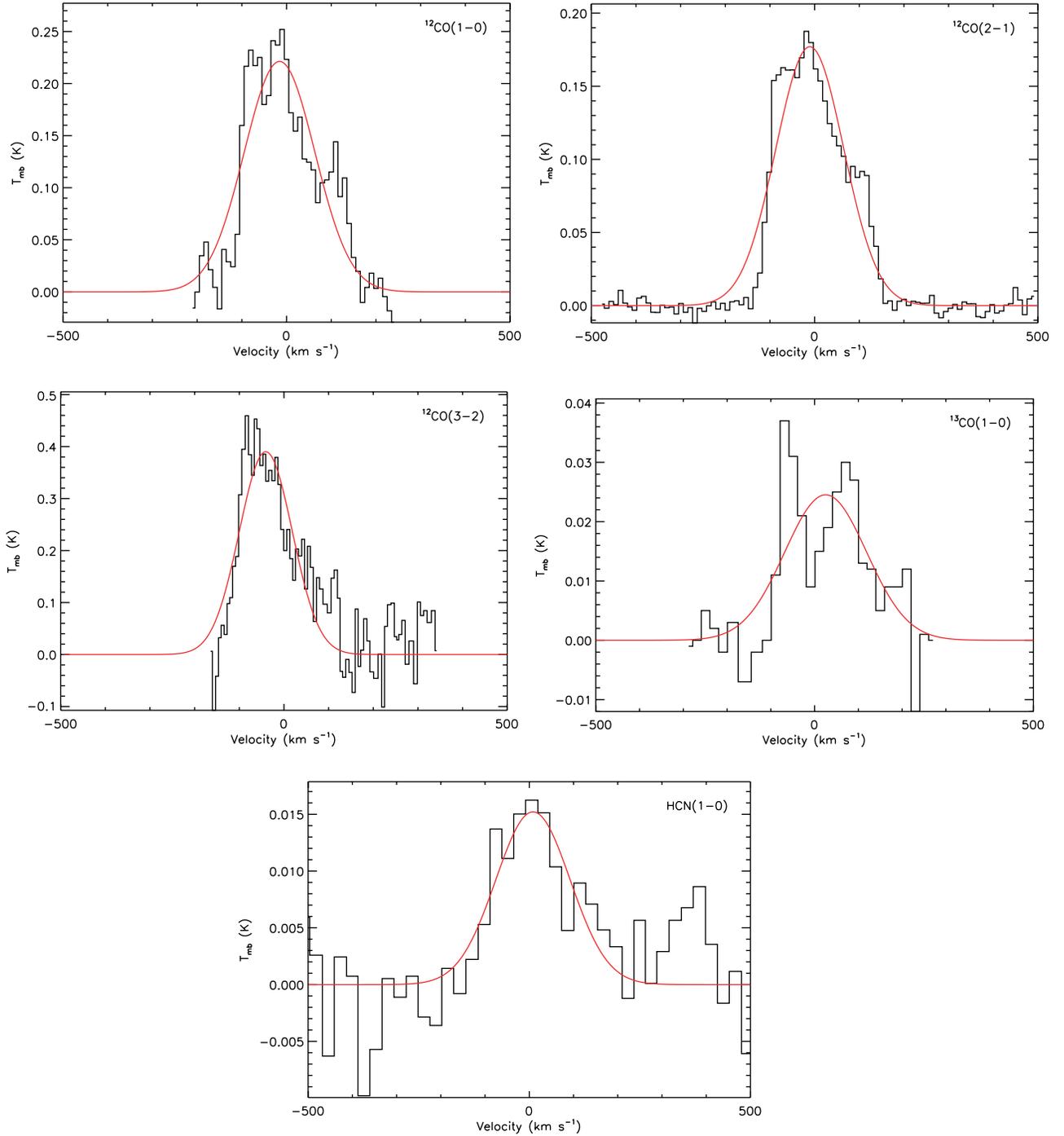

**Figure 2**. Integrated line profiles of the literature data detected at the center of NGC 2903. For each plot, the red line shows the best Gaussian fit to the spectrum. The name of the emission as listed in Tables 2 and 3 is indicated in the top-right corner of each plot. The rms on the spectrum is estimated using emission-free channels only.

$$f\left(v\right) = A \times e^{\frac{-(v-v_0)^2}{2\sigma^2}}, \tag{1}$$

where $A$ is the peak flux, $v_0$ is the central velocity, and $\sigma$ is the width of the profile edges of the Gaussian. The fits were carried out with the package MPFIT that employs a Levenberg–Marquardt minimization algorithm [24]. To prevent local minima in the parameter space, MPFIT was run several times for each spectrum with different initial guesses. The fitting parameters with the smallest $|1 - \chi^2|$ value were taken as the best fit.

For $^{13}$CO(1-0) and $^{12}$CO(3-2) spectra, we applied a beam correction to their integrated intensities by following the expression below [25–27]:

$$k = \left(\theta_{line}^2 + \theta_{source}^2\right) \left(\theta_{common}^2 + \theta_{source}^2\right), \quad (2)$$

where $\theta_{line}$ represents the original beam size of the observations (see Table 2), $\theta_{source}$ represents the linear size of the region studied, and $\theta_{common}$ is the common beam size of 28''. We defined the source size at the center of the galaxy using the highest spatial resolution map available. As no molecular map exists with such resolution, we used 24-µm images from the Spitzer Infrared Nearby Galaxy Survey [28]. We first considered a circular zone equal to the common beam size of 28'' in the center of the galaxy. We then defined the pixels with a flux above the 50% of the peak in that zone and calculated a total area including only those pixels. We finally took the source size as the diameter of a circle with the same area. The integrated intensities of $^{13}$CO(1-0) and $^{12}$CO(3-2) were multiplied by $k$ to obtain the intensities for the beam size of 28''. The integrated intensities are calculated for the common beam size for each transition and are listed in Table 3.

**Table 3**. Integrated intensities of the lines detected at the center of NGC 2903.

| Galaxy | $^{12}$CO(1-0) (K km/s) | $^{12}$CO(2-1) (K km/s) | $^{12}$CO(3-2) (K km/s) | $^{13}$CO(1-0) (K km/s) | HCN(1-0) (K km/s) |
|---|---|---|---|---|---|
| NGC 2903 | 43.86 ± 3.84 | 33.74 ± 1.51 | 37.16 ± 3.28 | 1.99 ± 0.45 | 3.21 ± 0.65 |

The beam-averaged total molecular gas mass in the unit of solar mass ($M_\odot$), $M_{H_2}$, at the center of NGC 2903 was estimated using the CO-to-$H_2$ conversion factor ($X_{CO}$), $^{12}$CO(1-0) integrated intensity, and the expression below:

$$\frac{M_{H_2}}{M_\odot} = 8.7 \times 10^5 \left(\frac{S_{1-0}}{K\,km\,s^{-1}}\right) \quad (3)$$

where $S_{1-0}$ is the integrated intensity of $^{12}$CO(1-0) in the unit of K km s$^{-1}$. The $X_{CO}$ value of $1.2 \times 10^{20}\,cm^{-2}(K\,km\,s^{-1})^{-1}$ was used in this study [29–31]. $X_{CO}$ is related to the beam-averaged $H_2$ column density ($N(H_2)$) as $X_{CO} = N(H_2) / S_{1-0}$. The values of $M_{H_2}$ and $N(H_2)$ enclosed with the adopted beam in the central region of NGC 2903 are listed in Table 4.

**Table 4**. The line ratios and molecular gas mass for NGC 2903.

| $R_{12}$ | $R_{13}$ | $R_{11}$ | $R_{D1}$ | $M_{H_2}$ ($10^7$ $M_\odot$) | $N(H_2)$ ($10^{21}$ cm$^{-2}$) |
|---|---|---|---|---|---|
| 1.30 ± 0.13 | 1.18 ± 0.15 | 22.00 ± 5.32 | 13.68 ± 3.04 | 3.80 ± 0.12 | 5.26 ± 0.46 |

### 3.3. Line ratios at the center

After obtaining the integrated line intensity for each transition at the center of the galaxy and applying the beam correction when necessary (i.e. by either convolving the cubes to the same beam size or estimating the beam correction factor), we took the ratios of the line intensities to study the physical conditions of the gas in the center. The ratios we obtained are $^{12}$CO(1-0) / $^{12}$CO(2-1) (hereafter $R_{12}$), $^{12}$CO(1-0) / $^{13}$CO(1-0) (hereafter $R_{11}$), $^{12}$CO(1-0) / $^{12}$CO(3-2) (hereafter $R_{13}$), and finally $^{12}$CO(1-0) / HCN(1-0) (hereafter $R_D$). The ratios are listed in Table 4.

### 3.4. Line ratios along the disc

We have the data cubes of $^{12}$CO(1-0) and $^{12}$CO(2-1) observations covering the entire disc of NGC 2903. This allows us to study the $R_{12}$ ratios along the disc. However, before taking the ratio of the data cubes we first need to convolve the cubes to a common beam size and make sure that the pixel size is also the same. We apply these corrections using *MIRIAD* tasks *colvol* and *regrid*. Both data cubes were convolved and grids were reconstructed accordingly, so that both cubes have the 2 × 2 arcsec$^2$ pixel size and a common beam size of 28''. We then created moment 0 maps for $^{12}$CO(1-0) and $^{12}$CO(2-1) as explained in Section 3.1. We finally took the ratio of the moment maps using the MIRIAD task *maths*. The resultant image is shown in Figure 3 and discussed in Section 5.2. We also compare our line ratios to those of literature data taken at the center of other galaxies (see Figure 4 and Section 5.2).

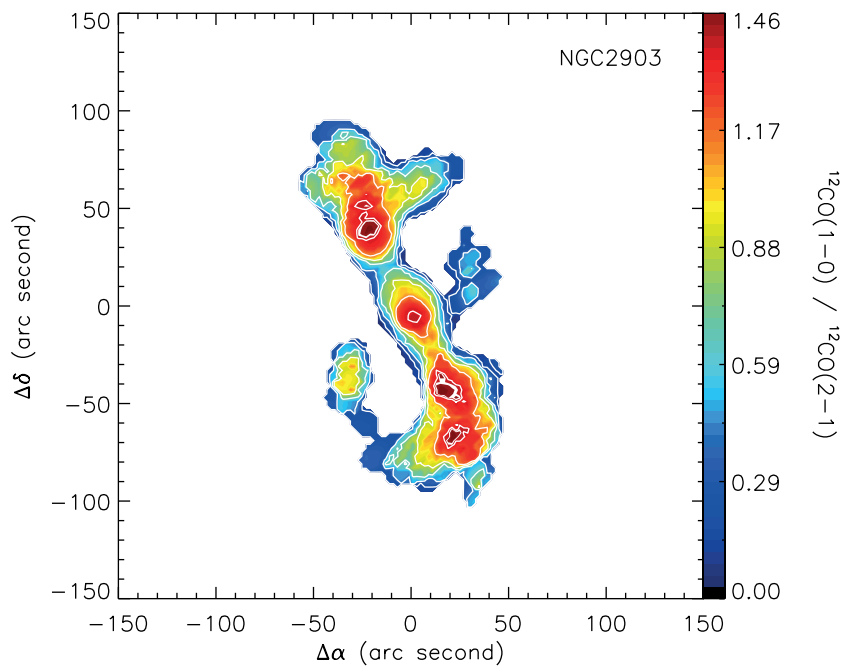

**Figure 3**. Contour maps for the ratio of $^{12}$CO(1-0) / $^{12}$CO(2-1), so-called $R_{12}$, along the disc of NGC 2903. North is up and east to the left in the image.

### 4. Modeling

We have multiple CO lines at the center of the galaxy, but only one dense gas tracer, namely HCN(1-0). We therefore run the model to characterize the physics of the CO gas only. However, the $R_D$ ratio will also be discussed in Section 5. We run the non-LTE radiative transfer code RADEX [15] to probe the physical conditions of the CO gas quantitatively. We use two methods to identify the best model representing the observed CO line ratios best: the $\chi^2$ (i.e. identifying the best-fitting model) and likelihood (i.e. identifying the most likely model). The input parameters for RADEX and the methods for the best model analysis are described in detail in the following subsections.

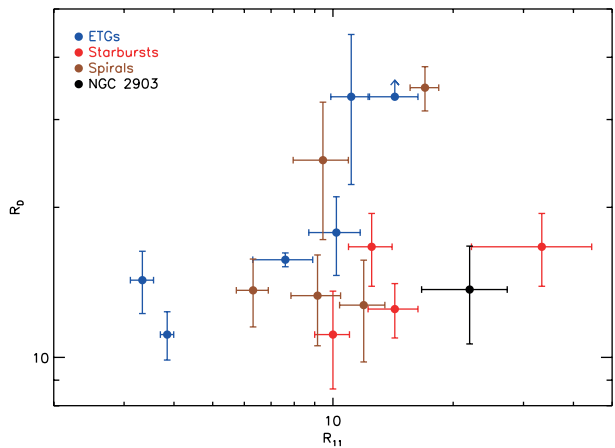

**Figure 4**. $R_{11}$ versus $R_D$ ratios in NGC 2903 and in selected galaxies from the literature. Red, blue, brown, and black filled circles represent starbursts [36], ETGs [37,38], spirals (unpublished data), and NGC 2903, respectively. Lower limit is indicated with arrow.

### 4.1. LVG code: RADEX

RADEX calculates the intensities of molecular lines considering excitation and deexcitation through the collisional processes with $H_2$, radiative processes, and the cosmic microwave background radiation ($T_{CMB} = 2.7$ K). The input parameters for RADEX are the kinetic temperature of the gas $T_K$, $H_2$ volume number density $n(H_2)$, CO column number density $N(CO)$, line width of the spectrum $\Delta v$ (namely the FWHM of the Gaussian fitted to the spectra), and, finally, intrinsic abundance ratio of $[^{12}C] / [^{13}C]$. We created model grids as follows. We kept $T_K$, $n(H_2)$, and $N(CO)$ as free parameters. The $T_K$ grid ranges from 10 K to 250 K (in steps of 5 K), $n(H_2)$ ranges from $10^2$ cm$^{-3}$ to $10^7$ cm$^{-3}$ (in steps of 0.25 dex), and $N(CO)$ ranges from $10^{13}$ cm$^{-2}$ to $10^{21}$ cm$^{-2}$ (in steps of 0.25 dex). This results in a parameter space containing a total of 33,957 model grids. We adopt a value of 70 for the intrinsic ratio of $[^{12}C] / [^{13}C]$, which is similar to the ratio seen in the center of our own galaxy Milky Way, in the local ISM, and in nearby starburst galaxies [32–35]. Although the average line widths of $^{12}CO$ and $^{13}CO$ are only slightly different, we took a line width of 170 km/s and 220 km/s for $^{12}CO$ and $^{13}CO$, respectively, instead of using one single average line width for the model. However, please note that the RADEX results are only minimally affected by the line widths (see [15]).

### 4.2. Best-fitting and most likely models

We followed two different approaches to define the best model parameters (i.e. $T_K$, $n(H_2)$, and $N(CO)$). First, we used the $\chi^2$ method to identify the best-fitting model. Second, we used a likelihood method to define the most likely model.

For our set of models, the $\chi^2$ is defined as:

$$\chi^2 \equiv \sum_i \left(\frac{R_{i,mod} - R_{i,obs}}{\Delta R_{i,obs}}\right)^2 \tag{4}$$

where $R_{mod}$ is the modeled line ratio and $R_{obs}$ is the observed line ratio with the uncertainty of $\Delta R_{obs}$. The model with the smallest reduced $\chi^2$ ($\chi^2_{r,min} = \chi^2/DOF$, where DOF represents corresponding degrees of freedom) is taken to be the best model. The contours of $\Delta\chi^2_r = \chi^2_r - \chi^2_{r,min}$ are shown in Figure 5. We also derived the most likely model from the probability distribution function (PDF) marginalized over the other two parameters. To do that, for each model parameter (i.e. $T_K$, $n(H_2)$, and $N(CO)$), we calculated the sum of the $e^{-\Delta\chi^2/2}$ terms for each possible value of the other two parameters. The PDFs are shown in Figure 6 and model results are listed in Table 5.

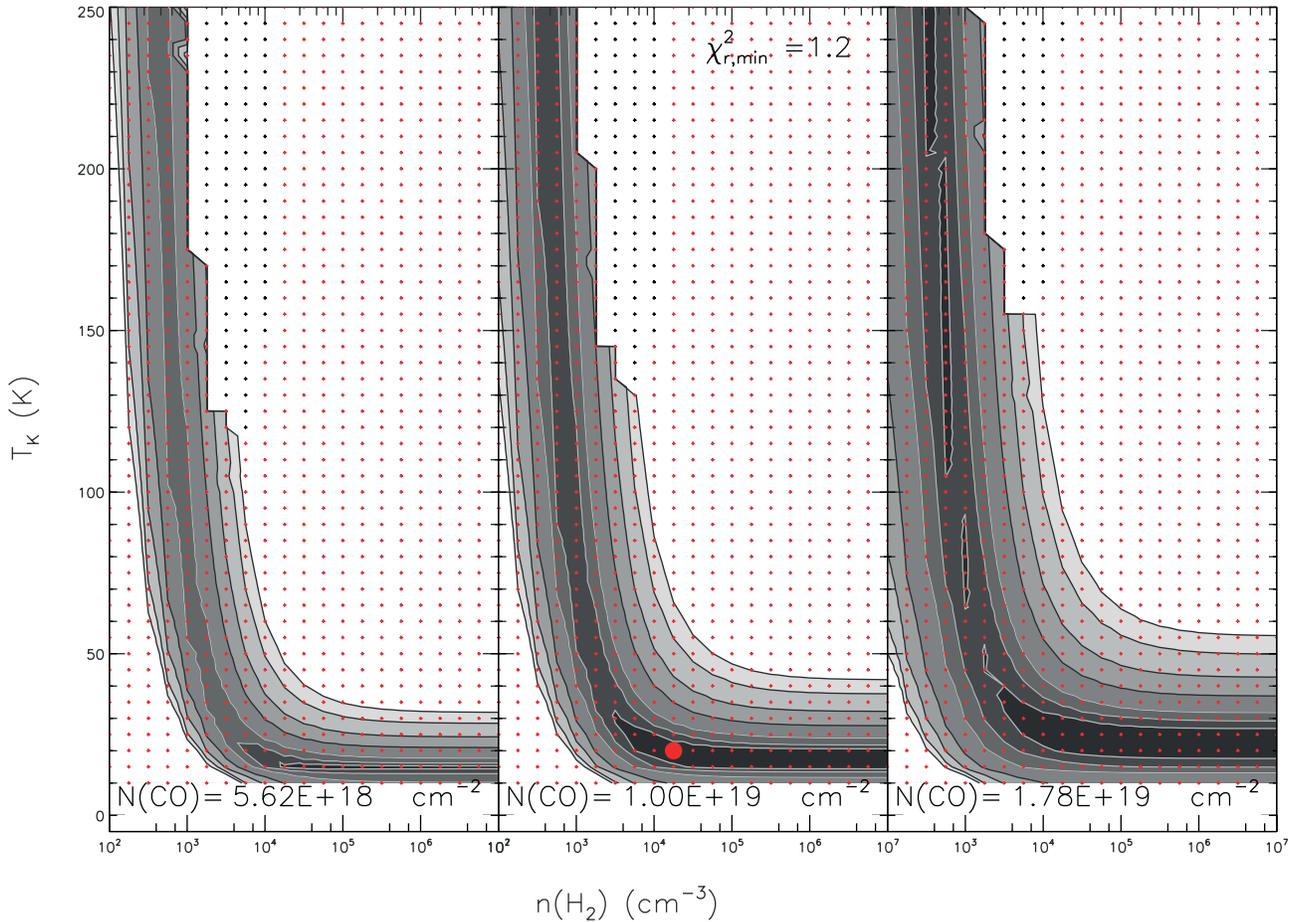

**Figure 5.** $\Delta\chi^2_r = \chi^2_r - \chi^2_{r,min}$ maps for CO gas in the center of NGC 2903. $\Delta\chi^2_r$ is shown as a function of $T_K$, and $n(H_2)$, for three values of CO column density $N(CO)$ centered around the best fit. The values of $N(CO)$ are indicated at the bottom of each panel and the values of reduced $\chi^2_{min}$ (namely $\chi^2_{r,min}$) at the top of the middle panel. The model outputs are indicated by red dots and the best fit model with a red filled circle. Black dots represent bad models (e.g., too low opacity; see [15]). The contour confidence levels are $0.2\sigma$ (16% probability that the best model is enclosed: the darkest gray-scale), $0.5\sigma$ (38%), $1\sigma$ (68%), $2\sigma$, $3\sigma$, $4\sigma$, and $5\sigma$ (the lightest gray-scale) for 3 degrees of freedom (three line ratios). The values for the level from $0.2\sigma$ to $0.5\sigma$ are 0.8, 1.8, 3.5, 8.0, 14.2, 22.1, and 28.0, respectively. The confidence levels from $0.2\sigma$ to $1\sigma$ are separated by gray lines and those from $2\sigma$ to $5\sigma$ by black lines.

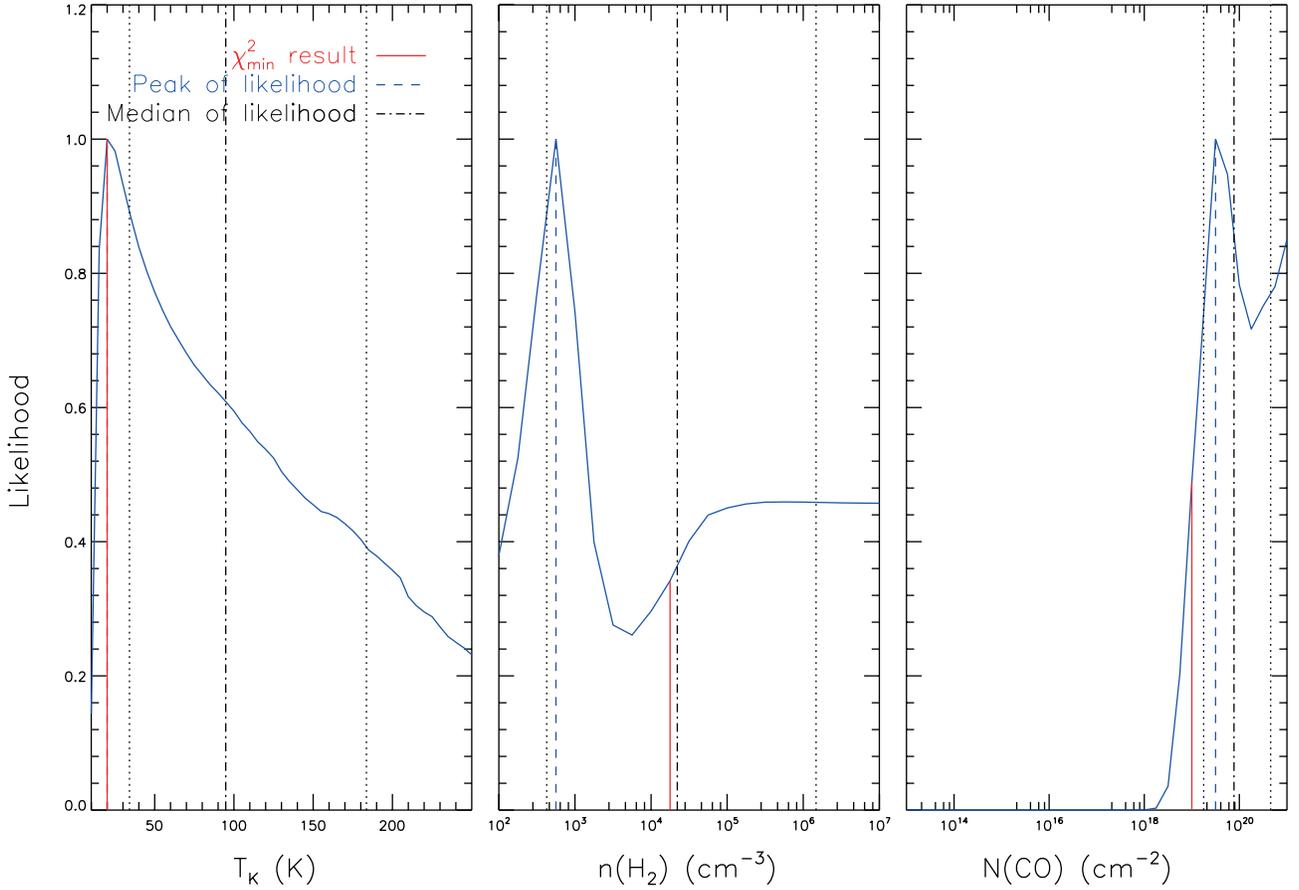

**Figure 6**. PDF of each model parameter, namely $T_K, n(H_2)$, and N(CO), marginalized over the other two. The peak (most likely) and median value within the model grid are identified with a dashed blue and dashed-dotted black line, respectively. The 68% ($1\sigma$) confidence level around the median is indicated by dotted black lines. The model results based on $\chi^2$ analysis are indicated by a solid red line in all panels.

**Table 5**. The best-fitting and most likely model results for NGC 2903.

| Parameter | $\chi^2$ | Likelihood |
|---|---|---|
| $T_K$ | 20 K | $95 \pm 74$ K |
| Log($n(H_2)$) | 4.2 cm$^{-3}$ | $4.3 \pm 1.7$ cm$^{-3}$ |
| Log($N(CO)$) | 19.0 cm$^{-2}$ | $19.9 \pm 0.7$ cm$^{-2}$ |

## 5. Results and discussion
### 5.1. Moment maps

As seen from panels a and b of Figure 1 (and also panels e and f), the gas is ubiquitous along the disc of NGC 2903. $^{12}$CO(2-1) is more extended than $^{12}$CO(1-0) through the disc. However, as seen from the same panels, $^{12}$CO(1-0) is brighter than $^{12}$CO(2-1). This could simply mean that the $^{12}$CO(2-1) observation is much deeper (i.e. higher signal-to-noise ratio) than the $^{12}$CO(1-0) observation, although it is fainter than $^{12}$CO(1-0) along most of the disc. We calculated the rms noise using emission-free channels in the cubes. The rms noises in the $^{12}$CO(1-0) and $^{12}$CO(2-1) data cubes are 21 mK and 6 mK, respectively, i.e. the $^{12}$CO(2-1) observation is indeed at least 3 times deeper than $^{12}$CO(1-0) observation.

As seen from the moment 1 maps (namely velocity maps) shown in Figure 1, panels c and d, NGC 2903 shows a regular velocity field, i.e. the gas is regularly rotating around the center of the galaxy. This indicates that the galaxy is not likely to be under any current/recent collisions and/or strong interaction with other galaxies. It is therefore not expected that such events could affect the observed line ratios.

## 5.2. Line ratios

$^{12}$CO(2-1) requires higher temperature than that of $^{12}$CO(1-0) to be excited and $^{12}$CO(3-2) requires a higher temperature than that of $^{12}$CO(2-1) and so on. $^{13}$CO gas is thinner than $^{12}$CO in galaxies. Therefore, ratios between the higher and lower transitions of the same molecule act like a thermometer for the gas cloud (e.g., the higher the $R_{12}$ ratio, the colder the CO gas and vice versa), while isotopic ratios are sensitive to opacity and column density (i.e. the higher the $R_{11}$ ratio, the thinner the CO gas).

We have $^{12}$CO(1-0) and $^{12}$CO(2-1) observations along the entire disc of NGC 2903. This allows us to study any changes in temperature through the disc of the galaxy. Distribution of the $R_{12}$ ratio along the disc is shown in Figure 3. As seen from Figure 3, the $R_{12}$ ratios are higher than 1 in most places along the disc, particularly at the center and at the edges of the disc on both sides. This indicates that $^{12}$CO(1-0) gas is dominant in the center and at the edges of the disc compared to $^{12}$CO(2-1), and the average temperature of the gas in these regions is therefore expected to be relatively less than that through the rest of the disc. On a different note, the high ratio seen at the edges (i.e. bright spots in northern and southern parts of the disc seen in Figure 3) could simply be because there is more gas along the line of sight.

The ratios of the integrated line intensities obtained at the center of the galaxy are listed in Table 3. As an indication of average kinetic temperature, the $R_{12}$ ratio has been obtained at the center of many galaxies in the literature. The $R_{12}$ ratio in the central region of 32 galaxies, including starbursts and interacting systems having higher star formation rates than spirals, like NGC 2903, was found to have a mean value of 1.1 [36], while the same ratios at the center of some ETGs are much higher than that of $1.30 \pm 0.13$ found at the center of NGC 2903 (see panel a of Figure 15 in [37]). This indicates that the gas at the center of NGC 2903 is warmer than that of ETGs and colder than the central gas complexes in starbursts and interacting systems.

We also compared our $R_D$ and $R_{11}$ ratios to those of a total of 15 galaxies selected from the literature including 4 starbursts and 1 spiral [38], 6 ETGs [37,39], and 4 spirals (unpublished data). The $R_D$ versus $R_{11}$ ratios of those galaxies together with our target galaxy NGC 2903 are shown in Figure 4. The $R_{11}$ ratio usually shows a radial gradient in spirals [40], although it can have local maxima in star-forming spiral arms [41]. As seen from Figure 4, NGC 2903 has the highest $R_{11}$ ratios among the sample (except one starburst, one spiral, and one ETG). This indicates that $^{12}$CO gas at the center of NGC 2903 is thinner than that seen in most of the comparison samples. Interestingly, the $R_{11}$ ratio found at the center of NGC 2903 ($R_{11} = 22.00 \pm 5.32$; see Table 4) is comparable (with a tendency of being higher) to the mean ratio of $13.6 \pm 6$ found at the center of 32 galaxies including mostly starbursts [36]. Stellar feedbacks (e.g., strong UV radiation from young massive stars and/or supernova explosions) in the center of NGC 2903 could be responsible for the gas becoming relatively more diffused. This indicates that current/recent star formation activity in the center of NGC 2903 is likely to be higher than most of the comparison samples and at least similar to that of starbursts. However, dense gas fraction, traced by CO-to-HCN ratio, should also be considered before making any firm conclusions on star formation activity.

HCN traces denser regions of the gas cloud than that of CO does. CO-to-HCN ratios thus trace the dense gas fraction within the star-forming gas clouds (i.e. the lower the R$_D$ ratio, the higher the dense gas fraction). As seen from Figure 4, NGC 2903 has a similar R$_D$ ratio (i.e. R$_D \leq 20$) compared to that of spirals, starbursts, and ETGs (except two ETGs and two spirals). This could simply be because of the selection effect: all the galaxies already have HCN emission detected in their centers, where the dense gas is generally concentrated in galaxies, and showing some level of star formation activity.

### 5.3. Modeling results

As revealed from Figure 5, there is degeneracy in $T_K$ due to its characteristic 'banana'-shaped $\Delta\chi_r^2$ contours or inverse $n(\text{H}_2) - T_K$ relationship ([25,37]; see also appendix C of [15]). The PDF of the kinetic temperature $T_K$ also shows a single-peaked distribution peaked at 20 K where the $\chi^2$ result meets with the peak, followed by a steady decrease. As seen from Table 5 and Figure 6, our model results indicate that H$_2$ volume number density $n(\text{H}_2)$ is the best-constrained parameter, i.e. the reduced $\chi^2$ and likelihood results agree within the error bars, and showing a single-peaked PDF followed by a flat distribution after about $n(\text{H}_2) = 10^5$ cm$^{-3}$. The same flatness is seen in the PDF of CO column number density $N(\text{CO})$ up to a value of $10^{18}$ cm$^{-2}$. The $1\sigma$ error bars around the median cover both the $\chi^2$ result and the peak for $n(\text{H}_2)$, while this is not the case for $T_K$ and $N(\text{CO})$ (except the peak).

We compare our best-fitting results (listed in Table 5) to that of literature studies done for the center and along the disc of one spiral and one ETG using similar lines and the same methods. Molecular gas properties along the disc of the spiral galaxy NGC 6946 and ETG NGC 4710 [25–27,37] were also studied using the LVG approximation. We found that $n(\text{H}_2)$ is to be higher in the center of NGC 2903 compared to that found in most of the regions studied in the arms of NGC 6946 [25] and in the center of NGC 6946 [26,27]. However, we found lower $T_K$ and higher $N(\text{CO})$ compared to those found in the center of NGC 6946 ($T_K = 40$ K and $T_K = 130$ K were found in the center of NGC 6946 for cold and hot gas components [26,27]). The central region of NGC 4710 was studied using higher resolution (about 6.5'') interferometric observations and a multiple gas component was found, i.e. cold-dense and hot-diffused gas components (see table B1 in [37]). Although the $T_K$ found in the center of NGC 2903 is an intermediate between hot-diffused and cold-dense components seen in NGC 4710, the gas is denser than both gas components located in the center of NGC 4710 and the $N(\text{CO})$ is similar. However, since the physical properties of the gas found in the center of NGC 2903 represent a beam-averaged quantity, i.e. averaged over a beam size of 28'', it is necessary to have a comparable angular resolution to that of NGC 4710 to make a better comparison.

All in all, the gas in the center of NGC 2903 is colder and denser than that in NGC 6946. This supports the inverse $n(\text{H}_2) - T_K$ relationship mentioned above, i.e. star formation is higher in NGC 6946, causing the temperature to be higher and the gas to be more diffused compared to the ISM in the center of NGC 2903. Interestingly, SFR of NGC 2903 (2.9 $M_\odot$ year$^{-1}$, see Table 1) is indeed lower than that of NGC 6946 (3.12 $M_\odot\, year^{-1}$, [42]) and it is more than 20 times higher than that of NGC 4710 (0.11 $M_\odot\, year^{-1}$, [43]).

### 6. Conclusions

In this study, we presented integrated line intensity ratios and radiative transfer modelling for the gas complexes in the center of NGC 2903. Our main conclusions are as follows:

i) CO gas is ubiquitous in NGC 2903 and $^{12}$CO(1-0) is brighter than $^{12}$CO(2-1) and $^{12}$CO(3-2) in the

center and along the disc, i.e. $R_{12} > 1$ and $R_{13} > 1$. The ratio of $^{12}CO(3\text{-}2)$ to $^{12}CO(2\text{-}1)$ is also higher than 1, i.e. $^{12}CO(3\text{-}2)$ is brighter than $^{12}CO(2\text{-}1)$. NGC 2903 shows a regular velocity field indicating no dominant effects of collisions to the observed line ratios.

ii) As an indication of gas kinetic temperature, the $R_{12}$ ratio in the center of NGC 2903 is lower than that found at the center of ETGs, while it is higher than that found at the center of starbursts and interacting systems. This indicates that the gas in the center of NGC 2903 is relatively warmer than that of ETGs and colder than that of starbursts.

iii) The $R_{11}$ ratio in the center of NGC 2903 is higher than that of the literature galaxy sample including 15 galaxies from three different galaxy types, i.e. ETGs, spirals, and starbursts. The dense gas fraction is, however, mostly similar to that of the comparison sample (except two ETGs and one spiral with higher $R_D$ values). Interestingly, the $R_{11}$ ratio seen at the center of the galaxy is also comparable (or even higher) to that seen in a large sample of starburst galaxies (32 galaxies in total), indicating a starburst-like environment and stellar feedbacks affecting the gas properties in the ISM of NGC 2903.

iv) Based on the LVG modeling, our best-fitting model results (in a $\chi^2$ sense) in the center of NGC 2903 are $T_K = 20$ K, $\log(n(H_2)) = 4.2$ cm$^{-3}$, $\log(N(CO)) = 19.0$ cm$^{-2}$. The most likely models results are $T_K = 95 \pm 74$ K, $\log(n(H_2)) = 4.3 \pm 1.7$ cm$^{-3}$, $\log(N(CO)) = 19.9 \pm 0.7$ cm$^{-2}$.

v) $n(H_2)$ is the best-constrained parameter, i.e. there is an agreement between the best-fitting and the most likely model results within the error bars, while the best-fitting model results for $T_K$ and $N(CO)$ are outside the range covered by the median of the PDFs. The flatness seen in the PDFs is responsible for the median to be shifted away from the best-fitting results and the peak. The model results indicate a higher $n(H_2)$ and intermediate $T_K$ compared to that of spiral galaxy NGC 6946 and ETG NGC 4710.

In summary, both methods, i.e. the line ratio diagnostics and modeling, suggest that the ISM in the center of NGC 2903 have an intermediate temperature and level of star formation activity (also supported by SFRs), thinner CO gas with similar dense gas fraction, and higher $H_2$ volume number density compared to that of spirals, ETGs, and starbursts. NGC 2903 is therefore a good laboratory to better understand the differences in star formation processes and their effects on the ISM seen in starburst galaxies with high SFRs and in ETGs, so-called red-and-dead systems. To achieve such a goal, multiple transitions of CO and dense gas tracers with higher resolution observations are needed.

**Acknowledgments**

The author would like to thank the anonymous referee for his/her insightful comments and suggestions. The author also acknowledges support from Van Yüzüncü Yıl University, Project Code FBA-2017-5874.